\documentclass{ws-p8-50x6-00}

\begin{document}

\title{Dispersion analysis for generalized spin polarizabilities}

\author{M. Vanderhaeghen, D. Drechsel, M. Gorchtein}

\address{Institut f\"ur Kernphysik, Johannes-Gutenberg-Universit\"at
  Mainz, D-55099 Mainz, Germany}

\author{A. Metz}

\address{CEA-Saclay, DAPNIA/SPhN, F-91191 Gif-sur-Yvette, France}

\author{B. Pasquini}

\address{ECT*, European Center for Theoretical Studies 
in Nuclear Physics and Related Areas, Villazzano (Trento), Italy}


\maketitle

\abstracts{We report on a dispersion relation formalism 
for the virtual Compton scattering (VCS) reaction on the proton,
which for the first time allows a dispersive evaluation of 4 generalized
polarizabilities. The dispersion formalism provides a new tool to analyze VCS
experiments above pion threshold, thus increasing the sensitivity to the
generalized polarizabilities of the nucleon.}

\section{Introduction}

Over the past years, the virtual Compton scattering (VCS) 
process on the proton, accessed through the $e p \to e p \gamma$
reaction, has become a powerful tool to provide new information on the internal
structure of the nucleon \cite{GuiVdh,Vdh00}. 
\newline
\indent
In the low energy regime below pion threshold, the outgoing photon in
the VCS process plays the role of a quasi-constant applied 
electromagnetic dipole field and, through electron scattering, 
one measures the spatial distribution of the nucleon response 
to this applied field. The response is
parametrized in terms of 6 generalized polarizabilities (GP's)
\cite{Gui95,Dre97}, which are functions of the square of the virtual 
photon four-momentum $Q^2$. 
The GP's provide valuable non-perturbative nucleon 
structure information, and have been calculated in different approaches. 
In particular, the GP's teach us about the interplay
between nucleon-core excitations and pion-cloud effects. 
\newline
\indent
The first dedicated VCS experiment has been
performed at MAMI \cite{Roc00} and two combinations 
of GP's have been determined at $Q^2$ = 0.33 GeV$^2$. 
Further VCS experiments are underway at lower $Q^2$ 
at MIT-Bates \cite{Sha97} and at higher $Q^2$ at JLab \cite{Bert93}.  
\newline
\indent
At present, VCS experiments at low outgoing photon energies 
are analyzed in terms of a low-energy expansion (LEX) as proposed 
by Guichon et al. \cite{Gui95}, 
assuming that the non-Born (i.e. nucleon excitation) response 
to the quasi-constant electromagnetic field is exclusively given 
by its leading term in the outgoing photon energy $\rm q'$. This term  
depends linearly on the GP's. 
As the sensitivity of the VCS cross sections to the GP's 
grows with the photon energy, it is
advantageous to go to higher photon energies, provided one can keep the
theoretical uncertainties under control when crossing the pion
threshold. The situation can be compared to real Compton
scattering (RCS), for which one uses a dispersion relation formalism
\cite{Lvo97,Dre00} to extract the polarizabilities at energies 
above pion threshold, with generally larger effects on the observables.  
We report here on recent work \cite{Pas00} to set up  
a dispersion formalism for VCS. This provides a tool to analyze 
VCS experiments at higher energies in order to extract 
the GP's from data over a larger energy range. 
It will be shown that the same formalism also provides for the first
time a dispersive evaluation of 4 GP's.

\section{Dispersion formalism for VCS}

The nucleon structure information obtained through VCS can be  
parametrized in terms of 12 non-Born invariant amplitudes, denoted by 
$F_i^{NB} (i = 1,...,12)$ \cite{Pas00}. 
The $F^{NB}_i$ are functions of 3 invariants for the
VCS process~: $Q^2$, $\nu = (s - u)/(4 M_N)$, and $t$ ($s$, $t$ and $u$ 
are the Mandelstam invariants for VCS, and $M_N$ denotes the
nucleon mass). 
\newline
\indent
Assuming analyticity and an appropriate high-energy behavior, 
the non-Born amplitudes $F^{NB}_i(Q^2, \nu, t)$ fulfill 
unsubtracted dispersion relations (DR's) with respect to the variable $\nu$ at
fixed $t$ and fixed virtuality $Q^2$~:
 \begin{equation}
{\mathrm Re} F_i^{NB}(Q^2, \nu, t) = 
{2 \over \pi}  {\mathcal P} \int_{\nu_{thr}}^{+ \infty} d\nu'  
{{\nu' \, {\mathrm Im}_s F_i(Q^2, \nu',t)} \over {\nu'^2 - \nu^2}},
\label{eq:unsub} 
\end{equation}
with ${\mathrm Im}_s F_i$ the discontinuities 
across the $s$-channel cuts of the VCS process. 
\newline
\indent
It has been shown \cite{Pas00} that 10 of the 12 invariant amplitudes
drop sufficiently fast at high energy and can 
be evaluated through unsubtracted dispersion integrals as in
Eq.~(\ref{eq:unsub}). The remaining
two amplitudes, denoted by $F_1$ and $F_5$, 
cannot be evaluated through an unsubtracted dispersion integral. 
This situation is similar to RCS, 
where 2 of the 6 invariant amplitudes cannot be evaluated 
by unsubtracted dispersion relations either \cite{Lvo97}.
\newline
\indent
The imaginary parts ${\mathrm Im}_s F_i$ 
in Eq.~(\ref{eq:unsub}) are calculated by use of unitarity. 
In our calculation, we saturate the dispersion integrals 
by the dominant contribution of the $\pi N$
intermediate states. For the pion photo- and electroproduction
helicity amplitudes, we use the MAID analysis \cite{maid00}, 
which contains both resonant and non-resonant pion production mechanisms.

\section{Dispersion results for generalized polarizabilities}

\begin{figure}[h]
\epsfxsize=7.4 cm
\epsfysize=7.4 cm
\centerline{\epsffile{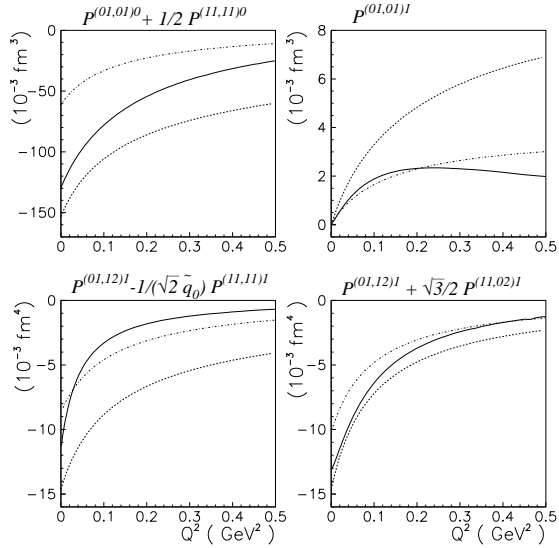}}
\vspace{-.45cm}
\caption[]{\small Dispersion results for 4 of the GP's of the proton 
(full curves), compared with results of $O(p^3)$ 
HBChPT \cite{Hem97} (dashed curves) and the linear $\sigma$-model
\cite{Met97} (dashed-dotted curves).}
\label{fig:polcomb}
\end{figure}

To obtain dispersive estimates for the GP's, one first 
expresses the GP's in terms of
the VCS amplitudes $F_i^{NB}$ at the point $\nu = 0$, $t = -Q^2$ at
finite $Q^2$, for which we introduce the shorthand~:
$\bar F_i(Q^2) \;\equiv\; F_i^{NB} \left(Q^2, \nu = 0, t = - Q^2 \right)$.
The relations between the GP's and the $\bar F_i(Q^2)$ can be found in
Ref.~\cite{Dre97}. 
\newline
\indent
Unsubtracted DR's for the GP's hold for those combinations of GP's 
that do not depend upon the amplitudes $\bar F_1$ and $\bar F_5$. 
Four such combinations of GP's 
have been found \cite{Pas00} and are shown in Fig.~\ref{fig:polcomb}~: 
one combination contains the scalar GP's, whereas the three other
combinations involve the spin GP's 
(see Ref.~\cite{GuiVdh} for the notations of the GP's).

The dispersion results are compared in Fig.~\ref{fig:polcomb} 
with the results of the 
$O(p^3)$ heavy-baryon chiral perturbation theory (HBChPT) \cite{Hem97}
and the linear $\sigma$-model \cite{Met97}. 
A comparison with those 
of HBChPT at $O(p^3)$ shows that a rather good agreement for  
$P^{\left(0 1, 1 2\right)1}$ + $\sqrt{3}/2 P^{\left(1 1, 0 2\right)1}$
is obtained,  
whereas for the GP's $P^{\left(0 1, 0 1\right)1}$ and 
$P^{\left(0 1, 1 2\right)1}$ - $1/(\sqrt{2} \, \tilde q_0) 
P^{\left(1 1, 1 1\right)1}$,
the dispersive results drop much faster with $Q^2$. This trend
is also seen in the relativistic linear $\sigma$-model, which 
takes account of some higher orders in the chiral expansion. 

\section{Dispersion results for VCS observables}

To construct the remaining two VCS amplitudes ($F_1$ and
$F_5$) in an unsubtracted dispersion framework, 
one can proceed in an analogous way as 
has been proposed by L'vov \cite{Lvo97} in the case of RCS.  
The unsubtracted dispersion integrals are firstly 
evaluated along the real $\nu$-axis in a finite range 
$-\nu_{max}\leq\nu\leq+\nu_{max}$ (with $\nu_{max}\approx$ 1.5~GeV).   
The remaining asymptotic contributions ($F_1^{as}$ and $F_5^{as}$)  
are then approximately parametrized by $t$-channel poles.
As a first step, we have parametrized $F_1^{as}$ through a $\sigma$-pole and 
$F_5^{as}$ through a $\pi^0$-pole, in analogy with the RCS case. 

\begin{figure}[h]
\epsfxsize=8. cm
\epsfysize=10. cm
\centerline{\epsffile{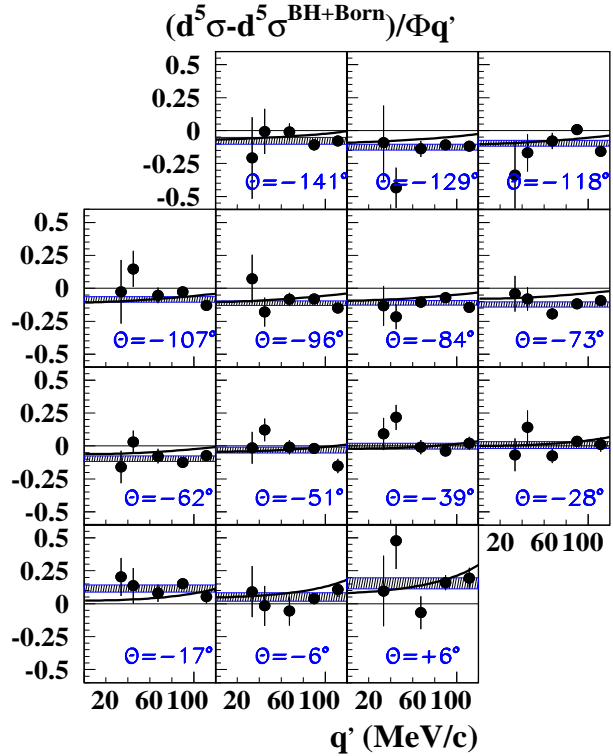}}
\vspace{-.4cm}
\caption[]{\small Photon energy dependence of the difference between 
the total VCS and Bethe-Heitler (BH) + Born cross sections, 
for different photon angles in the kinematics 
of the MAMI VCS experiment \cite{Roc00} 
: $Q^2 = 0.33$ GeV$^2$, $\varepsilon = 0.62$. 
The LEX gives a constant energy dependence, and is shown by
the band, which represents a fit to the data \cite{Roc00}.  
The dispersion calculation is represented by the solid curves.}  
\label{fig:qpdep}
\end{figure}
The resulting dispersive estimates are shown in Fig.~\ref{fig:qpdep}
and compared with the MAMI VCS data below pion threshold. It is seen
that in the low energy region (below pion threshold), the dispersive
results seem to support the LEX analysis, represented by the band in
Fig.~\ref{fig:qpdep}, from which two combinations of GP's have been
extracted in Ref.~\cite{Roc00}. It remains to be seen how more
refined parametrizations for the asymptotic contributions as well as
an estimate of $\pi\pi N$ intermediate states in the dispersion
integrals affect this result. 
\newline
\indent
The next step is then to apply the present VCS dispersion formalism to
higher energies in order to extract the nucleon GP's over a larger
range of energies from both unpolarized and polarized VCS data.

\section*{Acknowledgments}
This work was supported by the Deutsche
Forschungsgemeinschaft (SFB443), 
the EU/TMR Contract No. ERB FMRX-CT96-0008, and the ECT*.

\end{document}